\documentclass[11pt]{article}

\usepackage{graphicx}
\usepackage{dcolumn}
\usepackage{bm}

\usepackage[utf8]{inputenc}
\usepackage[T1]{fontenc}
\usepackage{geometry}
\usepackage{mathptmx}
\usepackage{etoolbox}

\usepackage{float}
\usepackage{tikz}
\usepackage{array}

\makeatletter
\def\@email#1#2{%
 \endgroup
 \patchcmd{\titleblock@produce}
  {\frontmatter@RRAPformat}
  {\frontmatter@RRAPformat{\produce@RRAP{*#1\href{mailto:#2}{#2}}}\frontmatter@RRAPformat}
  {}{}
}%
\makeatother
\begin{document}

\title{Crossing the Functional Desert:\\ Cascade-Driven Assembly and\\ Feasibility Transitions in Early Life}
\author{Galen J. Wilkerson\\
La Jolla, CA, USA\\
gjwilkerson@gmail.com}

\date{\today}

\maketitle
\begin{abstract}

The origin of life poses a problem of combinatorial feasibility: How can temporally supported functional organization arise in exponentially branching assembly spaces when unguided exploration behaves as a memoryless random walk? We show that nonlinear threshold-cascade dynamics in connected interaction networks provide a minimal, substrate-agnostic mechanism that can soften this obstruction. Below a critical connectivity threshold, cascades die out locally and structured input–output response mappings remain sparse and transient—a “functional desert” in which accumulation is dynamically unsupported. Near the critical percolation threshold, system-spanning cascades emerge, enabling discriminative functional responses. We illustrate this transition using a minimal toy model and generalize the argument to arbitrary networked systems. Also near criticality, cascades introduce finite-timescale structural and functional coherence, directional bias, and weak dynamical path-dependence into otherwise memoryless exploration, allowing biased accumulation. This connectivity-driven transition—functional percolation—requires only generic ingredients: interacting units, nonlinear thresholds, influence transmission, and non-zero coherence times. The mechanism does not explain specific biochemical pathways, but it identifies a necessary dynamical regime in which structured functional organization can emerge and be temporarily supported, providing a physical foundation for how combinatorial feasibility barriers can be crossed through network dynamics alone.  

\end{abstract}

\section{Introduction}

\begin{figure}[t]
\centering

\begin{tikzpicture}[>=latex, node distance=1.2cm, font=\small]

\node[circle, draw, minimum size=7mm] (A) at (0, 0.9) {$A$};
\node[circle, draw, minimum size=7mm] (B) at (0, -0.9) {$B$};
\node[circle, draw, minimum size=7mm] (C) at (2.2, 0) {$C$};

\node[anchor=west] at (2,.6) {$\phi$};

\draw[->, thick] (A) -- (C);
\draw[->, thick] (B) -- (C);

\node[anchor=west] (tbl) at (4.0, 0.55) {%
\renewcommand{\arraystretch}{1.2}%
\begin{tabular}{c c|c c}
$A$ & $B$ & $C$ (OR) & $C$ (AND) \\
    &     & $\phi \le 0.5$ & $\phi > 0.5$ \\
\hline
0 & 0 & 0 & 0 \\
0 & 1 & 1 & 0 \\
1 & 0 & 1 & 0 \\
1 & 1 & 1 & 1 \\
\end{tabular}%
};

\end{tikzpicture}

\caption{
Even a microscopic collection of interacting units can support functional accumulation.
Minimal interaction structure already supports non-trivial functional responses.
Two input nodes \(A\) and \(B\) (e.g., coarse-grained stimuli or activation sites) influence a downstream node \(C\) through directed couplings.
Node \(C\) is a thresholded unit with threshold parameter \(\phi\): It activates if the weighted fraction of active inputs exceeds \(\phi\).
For \(\phi \le 0.5\), activation of either input suffices and \(C\) implements an OR-like response.
For \(\phi > 0.5\), both inputs are required and \(C\) implements an AND-like response.
Because the interaction links and response rule remain coherent over the cascade timescale, structured multi-input responses can recur over short coherence intervals within the same dynamical regime, supporting a minimal form of functional accumulation.
}
\label{fig:toy_ltm_and_or}
\end{figure}
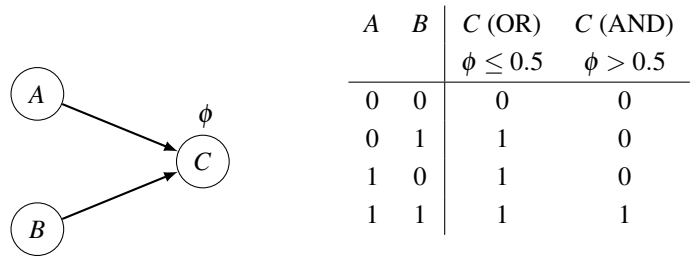


A central challenge in the origin of life is \textit{the emergence of functionally coherent organization over non-zero timescales in vast combinatorial spaces}. Previous work has shown that unguided exploration behaves effectively as a memoryless random walk, making reliable accumulation dynamically unsupported rather than merely improbable \cite{Eigen1971,endres2025}. 
In such spaces, the number of possible molecular or functional configurations grows super-exponentially with system size, rendering reliable accumulation of organized structure vanishingly unlikely without bias or mechanisms that preserve coherence in exploration over finite timescales.   

Although motivated by origin-of-life feasibility, the mechanism is stated for generic interacting units (molecules, proteins, genes, cells, neurons, agents), and the conclusions concern a substrate-agnostic transition in the space of realizable input–output response mappings.

Here, \textit{coherence} refers to the finite duration over which (i) the effective interaction structure remains sufficient for information transmission and (ii) the local response rules remain well-defined, so that a cascade can propagate and implement a reproducible input–output mapping over that timescale.  \textit{Accumulation} refers not to the growth of specific molecular objects, but to a more generalized notion: \textit{The emergence and finite-timescale support of structured functional organization, as revealed through non-trivial system responses to environmental inputs.} In this sense, our usage aligns with Schrödinger’s qualitative characterization of life as maintaining and amplifying order in far-from-equilibrium conditions.
Schrödinger’s emphasis on order referred not to formal entropy reduction, but to the finite-timescale maintenance of organized, functional structure in far-from-equilibrium systems \cite{schrodinger1967}.

Analyses across multiple frameworks suggest that unguided exploration in exponentially branching spaces fails catastrophically.  
Eigen’s error-threshold argument shows that faithful information accumulation becomes impossible beyond a critical mutation rate \cite{Eigen1971}.  
Endres has emphasized that even with unlimited time and energy, random exploration of chemical configuration space remains exponentially suppressed, creating a 'feasibility desert' \cite{endres2025}.  
Related feasibility bounds in origin-of-life models highlight similar obstructions, in which combinatorial explosion overwhelms selection, replication, or energetic driving alone \cite{sharma2023assembly}.  
These results collectively indicate that the central difficulty is not merely low probability, but a structural obstruction to accumulation itself.

Accordingly, the problem is best framed as one of \emph{feasibility} rather than improbability.  
The question is not whether life is unlikely, but whether any physical process exists that can make reliable accumulation possible in such vast configuration spaces at all.

Across biological organization, however, functional behavior is routinely mediated by thresholded, propagating processes that take the form of cascades.  
Gene regulatory networks exhibit switch-like activation and recruitment across interacting loci \cite{kauffman1969homeostasis}.  
Metabolic pathways transmit perturbations through coupled reaction chains.  
Neural systems display avalanche-like activity spanning wide spatial and temporal scales \cite{beggs2003neuronal}.  
Immune responses amplify local detection events into coordinated system-level action.  
Ecological systems propagate local disturbances through food webs and interaction networks \cite{may1972stability,parrish1999complexity}.  
In each case, local activation can either die out, propagate selectively, or spread indiscriminately depending on interaction structure and coupling.

Such cascade-like processes appear across biological scales, from molecular regulation to ecosystem dynamics, suggesting a form of universality in how biological systems transmit, amplify, and coordinate activity.  
Despite vast differences in substrate and scale, many living systems rely on thresholded propagation to organize responses to environmental perturbations.

Crucially, these cascades do not merely transmit signals.  
They often realize \textit{structured, non-trivial functional responses to environmental inputs, integrating multiple signals into coherent system-level behavior.}  
Examples include combinatorial gene regulation, multi-input neural decision-making, immune recognition, and coordinated collective motion.  
In each case, cascades enable responses that are neither random nor purely reactive, but instead exhibit functional structure.

Thus, while prior work has emphasized the infeasibility of assembling complex structures in vast configuration spaces, the present framework makes clear that the deeper obstruction lies in function space:  Below a critical interaction regime, structured input–output mappings cannot remain coherent over non-zero timescales.
Therefore the above 'feasibility desert' becomes a 'functional desert', in which meaningful functional organization is dynamically unsupported.

This motivates the central question of the present work:  
\emph{Can cascade dynamics provide a minimal physical mechanism that relieves the combinatorial feasibility obstruction?}  
Specifically, we ask whether connectivity-driven transitions in cascade behavior can enable the emergence of structured functional responses that recur over finite coherence timescales in otherwise unguided systems, thereby allowing accumulation in exponentially branching spaces.

Our goal is not to propose a biochemical pathway for abiogenesis, but to identify a general dynamical mechanism—independent of specific substrates—that can render accumulation feasible in principle.

\section{The Cascade Mechanism}

In this work, a \emph{cascade} refers to a thresholded spreading process in which local activations propagate through a network via neighbor-to-neighbor interactions.  
Nodes transition between coarse-grained states (e.g., inactive/active) based on whether incoming influence exceeds a specified threshold, allowing perturbations to either die out, spread selectively, or propagate system-wide.  We use the Linear Threshold Model (LTM) as a minimal, well-studied instantiation \cite{Granovetter1978, Watts2002, wilkerson_srep}.

The minimal ingredients required for cascades are generic and physically natural:  
(i) a network of interacting units,  
(ii) a threshold or nonlinearity governing state transitions,  
(iii) a mechanism for transmitting influence between units, and  
(iv) sufficient energetic or dynamical capacity to support state changes.  
No specialized biological machinery is required—only basic interaction and response rules.

Cascade dynamics arise generically in networked systems across physics, biology, and social systems.  
Thresholded propagation has been observed in neuronal avalanches, gene regulation, metabolic signaling, immune activation, ecological disturbance spreading, and collective behavior \cite{beggs2003neuronal,kauffman1993,parrish1999complexity}.  
The same qualitative behavior also appears in abstract network models, where simple local rules generate large-scale spreading phenomena \cite{Watts2002}.

Importantly, cascades occur even in structurally minimal networks.  
Random graphs represent a \emph{worst-case} scenario for organized propagation, lacking modularity, geometry, or functional specialization.  
Yet such networks still exhibit percolation transitions and cascade regimes once connectivity exceeds a critical threshold.  
This implies that cascade dynamics do not depend on fine-tuned structure or biological optimization.

Because cascades emerge from generic interaction rules, they do not require evolutionary design or biochemical specificity.  
Any system with sufficient connectivity, thresholded response, and transmission capacity can support cascade propagation, making the mechanism substrate-agnostic and broadly applicable.

Detailed treatments of cascade dynamics and their critical transitions can be found in prior work on threshold models \cite{Watts2002}, self-organized criticality \cite{bak1987soc}, and functional percolation in random networks \cite{wilkerson_fp}.  
Here, we focus on the conceptual role cascades play in enabling structured responses and accumulation, rather than on their specific microscopic implementations.

\section{From Connectivity to Functional Accumulation}
\subsection{Toy Example: Minimal Functional Cascades}

Imagine three system elements—molecules, neurons, cells, or other interacting units—that come together randomly and establish effective channels of influence among them. 
In physical systems, such interactions are typically mediated by contact or proximity, so small multi-unit encounters need not be implausible across many domains.
If these interactions are sufficient to form a minimally connected component, then activity at one element can, in principle, reach the others. In this informal sense, the system has crossed a basic connectivity threshold analogous to percolation: Local interactions are no longer isolated, but linked into a coherent interaction structure.

Once such connectivity exists, simple nonlinear response rules allow perturbations to propagate through the system. A minimal example is provided by a thresholded activation rule, in which a node becomes active only when the combined influence of its neighbors exceeds a fixed threshold. This type of rule is widely studied in threshold cascade models, including the Linear Threshold Model (LTM), and captures the essential feature that responses depend on collective input rather than purely additive effects \cite{Granovetter1978,Watts2002}.

In the toy network illustrated in Fig.~\ref{fig:toy_ltm_and_or}, two input nodes (A and B) influence a downstream node (C). The response of C depends on a single threshold parameter, denoted by $\phi$. When $\phi$ is low, activation of either A or B is sufficient to trigger C, yielding an OR-like response. When $\phi$ is higher, both A and B must be active to trigger C, yielding an AND-like response. The corresponding truth table makes this explicit: The same minimal structure supports distinct, nontrivial multi-input response patterns depending only on a simple local threshold.

This example demonstrates that \textit{even extremely few interactions can implement structured functional responses}. No specialized biological machinery is required; only connectivity and a nonlinear response rule are sufficient. The system is already performing a simple form of Boolean computation, integrating multiple inputs into an organized output pattern, consistent with earlier demonstrations of computation emerging in network cascades \cite{wilkerson_srep}.

Importantly, such functional responses require only finite-timescale coherence. The interaction links must remain intact long enough for influence to propagate, and the response rule at C must remain coherent long enough for inputs to be integrated. No long-term memory, storage, or repeatability is assumed. The coherence required is strictly local in time and limited to the duration of the cascade itself.

In this sense, even a tiny connected system can begin to "cross the functional desert." Rather than requiring the accumulation of specific molecular objects, the system accumulates functional organization: Structured input--output relationships that can, in principle, be reused and extended within recurring coherence intervals. The toy model provides a concrete, intuitive demonstration of how minimal interactions can already support nontrivial functional structure.


\subsection{General Mechanism: Cascades and Finite-timescale coherence}

The toy example illustrates a more general mechanism that applies to networked systems of arbitrary size. Consider a system composed of many interacting nodes. As connectivity increases, the system undergoes a percolation transition: A giant connected component emerges, allowing influences to propagate across large portions of the network \cite{Watts2002}. Random graph models provide a conservative baseline for this transition, demonstrating that cascade-capable connectivity does not require specialized architecture or biological optimization.

As noted in the Introduction, thresholded response rules in connected systems generically give rise to such cascades across a wide range of biological and social contexts.  Cascades arise generically in connected networks with nonlinear response rules and do not require biological fine-tuning or specialized architectures.

For a cascade to occur, as we have  just seen above, the interaction structure must persist long enough for influence to propagate. The connectivity and susceptibility profile of the network must remain coherent over the cascade timescale.

Similarly, for a system to exhibit a functional response—i.e., a structured mapping from inputs to outputs—the response rules at the nodes must remain coherent over the same timescale.  

Under these minimal conditions, cascades can integrate multiple inputs into organized system-level responses. 
Because physical interactions are proximity-mediated, small interaction motifs and short-lived channels of influence need not be implausible in many domains; however, the central requirement here is simply that connectivity and response rules remain coherent over the cascade timescale.
\textit{In sufficiently connected systems, this is enough for structured input--output response mappings to be realized and to recur over short coherence intervals, without requiring long-term memory, stored programs, or a single combinatorial leap to highly organized assemblies.}
Functional organization thus becomes accessible directly from connectivity and nonlinear dynamics whenever influence can propagate and thresholds implement multi-input integration.
This connectivity-driven transition in the accessibility of functional responses, coinciding with cascade emergence near percolation criticality, is what we refer to as \textit{functional percolation}, and has been studied elsewhere \cite{Watts2002,wilkerson_srep,wilkerson_fp}.

This mechanism has direct implications for the problem of combinatorial feasibility in origin-of-life scenarios. By enabling structured, biased responses to environmental inputs through transient but coherent interaction dynamics, cascades provide a route to organized functional accumulation even in the absence of long-term memory or replication.

\section{Consequences of the General Mechanism}

\subsection{Finite-timescale coherence}

The core argument implies two minimal and distinct forms of coherence. 
\emph{Structural coherence} refers to the coherence of interaction connectivity and susceptibility profiles over the timescale of a cascade. 
For influence to propagate, the network structure must remain intact long enough for signals to traverse multiple nodes \cite{Watts2002,Granovetter1978}.

\emph{Functional coherence} refers to the coherence of response mappings over the realization timescale. 
For a system to exhibit a structured functional response, the rules governing how nodes integrate inputs must remain stable long enough for inputs to be transformed into organized outputs \cite{wilkerson_srep,shalizi_crutchfield2001computational}.

Crucially, neither form of coherence assumes long-term stability, stored memory, inheritance, or replication. 
Coherence here means only that relevant structures remain coherent over the duration of a single realization. 
This minimal notion of coherence is sufficient to support cascades and functional responses without invoking biological machinery or evolutionary processes \cite{wilkerson_fp}.

\subsection{Accumulation}

We reiterate that \textit{accumulation}, in this framework, does not refer to the growth of specific molecular objects or material structures. 
Instead, it refers to biased exploration across realizations, in which structured functional organization is preserved rather than erased.

In purely memoryless exploration of combinatorial spaces, partial progress is lost between realizations, preventing reliable buildup of organization, much like the "monkeys typing Shakespeare" intuition \cite{borel1913hasard}. 
Finite-timescale structural and functional coherence prevents this total erasure by allowing coherent response patterns to recur across perturbations.

As a result, systems can accumulate organized functional distinctions even without long-term memory or replication. 
This form of accumulation directly addresses feasibility constraints in exponentially branching spaces, where unguided random search would otherwise fail \cite{Eigen1971,Orgel2008,sharma2023assembly,endres2025}.

\subsection{Directionality and Bias}

In this framework, directionality refers not to the direction of signal propagation per se, but to the \emph{directionality of accumulation} in functional and assembly space \cite{endres2025}.  
Cascades, together with minimal structural and functional coherence, 
bias which functional organizations are realized and can recur over short coherence timescales.

Because node responses depend on collective inputs, certain response patterns recur more readily than others.  
This introduces a bias in how structured functional organizations are explored and supported for recurrence over finite coherence times, even in the absence of design, optimization, or selection \cite{Watts2002,schreiber2000transfer}.

As a result, accumulation becomes weakly path-dependent.  Here, path-dependence refers to the fact that functional responses depend on the specific dynamical routes by which perturbations propagate through the interaction network.  
This dependence arises from nonlinear threshold dynamics and finite coherence times, and does not imply memory, learning, or structural modification across realizations.
The system’s exploration of its combinatorial space is therefore no longer symmetric or memoryless, but directionally constrained toward particular organizations \cite{shalizi_crutchfield2001computational}.

Importantly, this directionality is non-teleological.  
It does not reflect purpose, intention, or evolutionary selection, but arises as a structural consequence of cascades operating in finite-timescale coherent interaction networks \cite{kauffman1993}.

\section{Feasibility and Crossing the Functional Desert}

\begin{figure*}[t]
    \centering
    \includegraphics[width=0.6\linewidth]{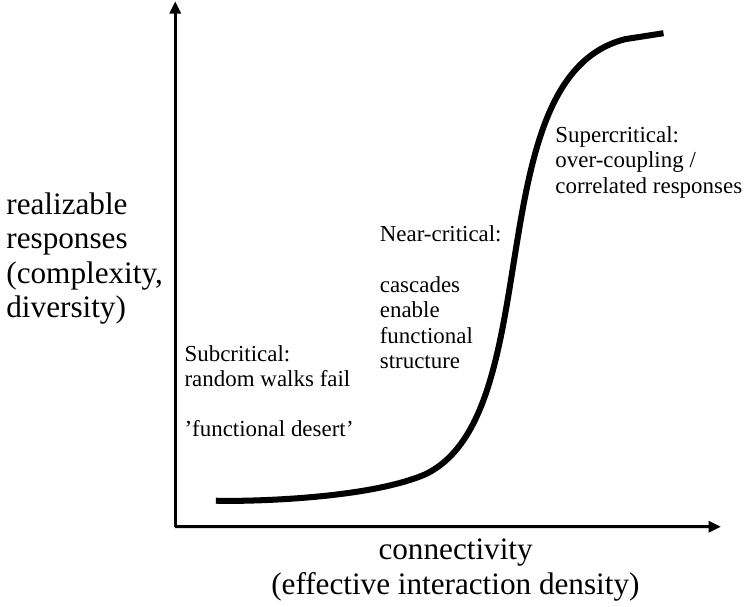}
    \caption{
Functional percolation as a feasibility transition in function space (schematic).
As effective interaction connectivity increases, threshold-cascade dynamics organize into three regimes.
Below the transition (subcritical), cascades die out locally and realizable input--output response functions are sparse and nonpersistent.
Near the functional percolation threshold, non-saturating system-spanning cascades emerge, sharply expanding the accessible repertoire of response functions that recur over finite coherence timescales and enabling long-range directed influence.
Above the transition (supercritical), increasing connectivity strengthens collective coupling, reducing functional diversity and constraining responses toward more correlated global patterns without eliminating information flow.
The near-critical regime constitutes a narrow feasible window in which accumulation in an exponentially branching space becomes possible through finite-timescale coherence and bias, without collapse or saturation of functional distinctions.
This schematic summarizes the qualitative regime structure observed in quantitative threshold-cascade and functional percolation simulations near criticality; see Refs. \cite{wilkerson_fp,wilkerson_srep} for numerical results.
}
    \label{fig:schematic_fp}
\end{figure*}

As discussed in the Introduction, origin-of-life scenarios face a fundamental problem of \emph{combinatorial feasibility}.  
In exponentially branching assembly spaces, unguided exploration behaves as a memoryless random walk, making reliable accumulation of organized structure dynamically unsupported rather than merely improbable.  
Eigen’s error-threshold argument \cite{Eigen1971}, Endres’ feasibility bounds \cite{endres2025}, and assembly-theoretic constraints \cite{sharma2023assembly} all point to the same conclusion: Random exploration cannot sustain accumulation.

The preceding sections identified a qualitative transition that occurs as interaction connectivity increases and cascades become system-spanning.  
This transition, which we refer to as \emph{functional percolation} \cite{Watts2002,wilkerson_srep,wilkerson_fp}, marks a change in the types of functional responses a system can reliably realize and retain.

Below the functional percolation threshold, cascades die out locally.  
Input--output response mappings are sparse, transient, and nonpersistent, preventing the accumulation of structured functional organization.  This is the 'functional desert'.

Above the threshold, at high connectivity, cascades become increasingly dominated by collective coupling.  
Many distinct inputs trigger more strongly correlated global responses, reducing functional diversity through over-constrained dynamics rather than eliminating information flow.

Near the critical transition, however, cascades are system-spanning without being fully saturating.  
This regime supports maximal functional discrimination, long-range directed influence, and the emergence of a rich repertoire of response mappings that recur over finite coherence intervals [Fig.~\ref{fig:schematic_fp}] \cite{wilkerson_fp,wilkerson_srep}.  
Only in this narrow window does structured functional organization become both realizable and capable of recurrence across realizations.  An 'oasis' in the functional desert.

Thus, near criticality, cascades bias which functional organizations are realized.  
Exploration becomes directionally constrained in functional and assembly space, introducing path-dependence in what is built \cite{endres2025,shalizi_crutchfield2001computational}.  
Accumulation is no longer symmetric or transient, but structured and supported for recurrence over finite coherence timescales.
Quantitative characterization of this weak dynamical path-dependence in specific models is left for future work.

In this sense, functional percolation constitutes a \emph{feasibility transition}.  
Random exploration in combinatorial spaces is converted into directionally biased accumulation through the emergence of discriminative cascades that remain coherent over finite timescales \cite{endres2025,shalizi_crutchfield2001computational}.  
Accumulation becomes physically supported, not merely statistically possible.

The functional desert is therefore crossed not by improbability reduction, but by a change in the dynamical regime governing exploration and retention of functional organization.

\section{Discussion}

\subsection*{From Function Space Back to Physical Systems}

Throughout this work, feasibility has been framed in terms of abstract function space: the space of realizable input--output response mappings supported by interacting systems.  
These mappings, however, correspond to concrete, observable behaviors in physical substrates.

In chemical systems, structured functional responses can arise when reaction pathways activate only under specific combinations of conditions, such as the co-presence of multiple reactants or catalysts \cite{wachtershauser1990evolution}.  
In gene regulatory networks, thresholded transcriptional responses integrate multiple regulatory signals to produce specific expression patterns \cite{kauffman1969homeostasis}.  
In neural systems, population-level cascades integrate sensory inputs into coherent activity patterns, with near-critical dynamics supporting rich, discriminative responses \cite{beggs2003neuronal}.  
In ecological networks, perturbations can propagate through species interactions, producing coordinated population-level responses \cite{may1972stability,parrish1999complexity}.

Across these domains, structured functional responses to inputs correspond to collective behaviors with non-zero coherence over cascade/interaction timescales, enabled by interaction topology, response thresholds, and timescale coherence rather than by detailed biochemical specialization.
Random graph models provide a conservative baseline for this transition, demonstrating that cascade-enabled functional responses do not rely on specialized structure or biological optimization.

More generally, the bridge from microscopic interaction to macroscopic organization is provided by compositional structure.
If small interaction motifs can implement structured responses over their coherence timescales, then those motifs can function as modules that participate in larger-scale networks of influence.
As interaction scope increases---through additional contacts, proximity-mediated encounters, or higher effective connectivity---modules can combine to form richer functional organizations without requiring that large assemblies appear all at once.
This perspective applies beyond biochemistry: neuronal circuits, cellular signaling modules, ecological subsystems, and social groups all exhibit local interaction motifs that can combine into higher-order organizations.
Accordingly, the present framework connects minimal cascade-enabled function to the scalable emergence of organized systems by emphasizing modular composition rather than abrupt large-scale assembly.

Empirical assembly-theory studies emphasize that real chemical systems occupy a highly structured and historically constrained region of object space, characterized by reuse of substructures and non-random assembly pathways \cite{sharma2023assembly}.
Functional percolation offers a dynamical mechanism through which such biased exploration and partial finite-timescale coherence can emerge from generic interaction networks, without requiring explicit replication or selection.  Whether analogous cascade-driven regimes existed in specific prebiotic chemistries remains an open empirical question.

As shown in Secs.~2--4, functional percolation can be interpreted as a transition in physical interaction networks.  
When effective connectivity, response thresholds, and coherence times jointly exceed minimal requirements, cascades support discriminative response patterns that remain coherent over finite coherence timescales rather than isolated or indiscriminate activity.

Below this transition, perturbations remain local and response mappings are sparse and transient; near criticality, cascades are system-spanning yet discriminative; above it, over-coupling increasingly constrains responses, reducing functional diversity through correlated collective dynamics [Fig.~\ref{fig:schematic_fp}].

The essential requirement is therefore not biochemical specialization, but the presence of thresholded interactions operating over coherent timescales in physically connected systems.

\subsection*{Relation to Existing Feasibility Arguments}

Classical origin-of-life constraints emphasize the difficulty of accumulation in combinatorial spaces.

Eigen’s error-threshold argument identifies a fidelity bottleneck: Information cannot be reliably preserved when mutation rates exceed a critical limit \cite{Eigen1971}.  
Assembly theory highlights strong path-dependence and bias in the construction of complex objects, implying severe constraints on feasible assembly histories \cite{sharma2023assembly}.  
Endres has recently formalized feasibility in terms of a rate--distortion framework, showing that molecular persistence times and environmental entropy impose strict bounds on what can accumulate, even with unlimited time and energy \cite{endres2025}.

The present framework does not replace these constraints.  
Instead, it identifies a complementary, network-level mechanism by which coherence and bias can emerge \emph{before} replication, heredity, or high-fidelity information storage.  

Functional percolation introduces a mesoscopic coherence window: Structured functional responses can recur across realizations through cascade dynamics alone.  
This provides a dynamical route to biased accumulation in function space, even when microphysical persistence remains limited.

\subsection*{What This Framework Does Not Explain}

This work does not propose a specific chemical pathway to abiogenesis.  
It does not specify protocell architectures, metabolic cycles, genetic codes, or replication mechanisms.

Rather, it identifies a necessary \emph{dynamical feasibility transition}: A regime in which structured functional distinctions can be generated, retained, and biased across realizations.  
Such a transition is a prerequisite for any later biochemical or evolutionary elaboration, but it is not itself a model of life’s origin.

\subsection*{Implications for Experiments and Computation}

The framework suggests several testable predictions.

In experimental and computational systems, one should observe:

\begin{itemize}
\item Expansion of functional response repertoires near connectivity thresholds.
\item Maximal input--output discriminability near criticality.
\item Reduction of functional diversity under excessive coupling due to over-constrained collective dynamics.
\item Sensitivity of accumulation to interaction coherence times.
\end{itemize}

These predictions can be explored in synthetic gene networks, neural cultures, chemical reaction systems, and agent-based models.  
Empirically, assembly-theoretic datasets \cite{sharma2023assembly} and perturbation--response studies in biological networks provide natural testing grounds for the presence of biased functional accumulation.

\subsection*{Substrate-Agnostic Universality}

Finally, the mechanisms identified here belong to a broad universality class.  
They require only:

\begin{itemize}
\item Interacting units,
\item Thresholded response rules,
\item Transmission of influence,
\item Finite coherence times.
\end{itemize}

These ingredients are ubiquitous across physical, biological, and artificial systems.  
As such, functional percolation represents a substrate-agnostic route by which combinatorial feasibility barriers can be softened through network dynamics alone.

The emergence of accumulation, bias, and directionality does not require design, selection, or teleology.  
It arises as a structural consequence of cascades operating near critical connectivity.

\section{Conclusion}

The origin of life confronts a fundamental problem of \emph{combinatorial feasibility}: In exponentially branching assembly spaces, unguided exploration behaves as a memoryless process that cannot reliably support the accumulation of organized structure \cite{Eigen1971,endres2025,sharma2023assembly}.  
This obstruction is structural rather than probabilistic.  
The question is not whether life is unlikely, but whether any physical mechanism exists that can render accumulation dynamically feasible at all.

In this work, we have identified a minimal, substrate-agnostic mechanism by which such feasibility can emerge.  
Thresholded cascade dynamics in connected interaction networks give rise to a connectivity-driven transition—\emph{functional percolation}—in the accessibility and finite-timescale coherence of structured input–output response mappings.
Near this transition, cascades are system-spanning yet discriminative, enabling biased, partially persistent functional organization without requiring long-term memory, replication, or hereditary storage.

The mechanism relies only on generic physical ingredients: interacting units, nonlinear response thresholds, transmission of influence, and finite coherence times.  
No biochemical specialization, evolutionary optimization, or teleological structure is assumed.  
Coherence arises over the duration of cascades themselves, allowing functional responses to recur across realizations and introducing directionality and bias into exploration of functional and assembly space.

Importantly, the claims advanced here are conditional and non-teleological.  
Functional percolation does not explain the origin of specific biochemical systems, protocells, or genetic mechanisms.  
Rather, it identifies a necessary \emph{dynamical feasibility transition}: a regime in which structured functional distinctions can be generated, retained, and biased across realizations.  
Such a regime constitutes a prerequisite for any subsequent evolutionary or biochemical elaboration, but it is not itself a model of life’s origin.

This work highlights a general physical route by which combinatorial barriers to accumulation can be softened through network dynamics alone.  
Critical cascades provide a minimal, universal mechanism for the emergence of biased functional organization in otherwise unguided systems, offering a principled foundation for future theoretical, computational, and experimental investigations into the physical conditions that make life possible.

\section*{Acknowledgments}

 Acknowledgment of discussions with Henrik Jeldtoft Jensen, Imperial College London.

\bibliography{refs}
\bibliographystyle{acm}

\end{document}